\documentclass[prd,aps,floats,twocolumn,nofootinbib]{revtex4-1}
\usepackage{slashed}
\usepackage{mathtools}
\usepackage{amsfonts}
\usepackage{amssymb}
\usepackage{epsfig}

\begin{document}

\title{A contextual Planck parameter and the classical limit in quantum
cosmology}
\date{\today }
\author{John D. Barrow}
\affiliation{DAMTP, Centre for Mathematical Sciences, Cambridge University, Wilberforce
Rd., Cambridge CB3 0WA, United Kingdom}

\author{Jo\~{a}o Magueijo}
\email{j.magueijo@imperial.ac.uk}
\affiliation{Theoretical Physics Group, The Blackett Laboratory, Imperial College, Prince Consort Rd., London, SW7 2BZ, United Kingdom}

\begin{abstract}
\begin{center}
{\bf Abstract}
\end{center}
We propose that whatever quantity controls the Heisenberg uncertainty
relations (for a given complementary pair of observables) it should be
identified with an effective Planck parameter. With this definition it is
not difficult to find examples where the Planck parameter depends on the
region under study, varies in time, and even depends on which pair of
observables one focuses on. In quantum cosmology the effective
Planck parameter depends on the size of the comoving region under study, and
so depends on that chosen region and on time. With this criterion, the
classical limit is expected, not for regions larger than the
Planck length, $l_{P}$, but for those larger than $%
l_{Q}=(l_{P}^{2}H^{-1})^{1/3}$, where $H$ is the Hubble parameter. In
theories where the cosmological constant is dynamical, it is
possible for the latter to remain quantum even in contexts where everything
else is deemed classical. These results are derived from standard
quantization methods, but we also include more speculative cases where 
\textit{ad hoc} Planck parameters scale differently with the length scale
under observation. Even more speculatively, we examine the possibility that
similar complementary concepts affect thermodynamical variables, such as the
temperature and the entropy of a black hole. 
\end{abstract}

\maketitle

\newtheorem{conj}{Conjecture}





\section{Introduction}
Although promoting the constants of Nature to dynamic fields is hardly a
novelty, structural fundamental constants, such as $c$ and $\hbar $, are
usually spared this transformation. In particular, Planck's constant (with
notable exceptions, for example~\cite{APbook,lizzi,Budi1,Budi2}) has had its constancy left largely
unchallenged. This is perhaps less warranted by physics conservatism than
one might think. The true hallmark of quantum behaviour is the impossibility
of jointly measuring complementary observables, usually (but not always)
resulting from their non-commutativity and associated Heisenberg uncertainty
relations~\cite{heis,schrod,poulin,dave}. Whatever function controls these relations (or the commutator)
acts as an \textit{effective} Planck parameter, even if it is not a constant%
\footnote{%
We will adopt the term \textquotedblleft parameter\textquotedblright\ for
this reason.}, and even if $\hbar $ is only one of its contributing factors,
as we shall find in concrete examples in this paper.
Hence, it is possible
that the effective Planck parameter has a variability induced by context and
circumstance, depending on the system under study and the chosen pair of
variables, and possibly evolving in time for a given system and pair.

In this paper we make this point in a variety of situations. In Section~\ref{sec:MSS}
we consider standard quantum cosmology, but rather than extending the action to
a whole closed manifold, we consider the context of fixed comoving spatial regions. 
The effective Planck parameter is then found to depend on the size of the comoving region under study, and
so depends on the chosen region and on time. By studying this parameter we discover that 
the  classical limit is expected, not for regions larger than the
Planck length, $l_{P}$, but for those larger than 
a mesoscopic scale, of the order of the proton scale. 

In Section~\ref{sec:GB} we extend our considerations to theories
for which the cosmological constant, $\Lambda$,  is dynamical, with the surprising result that it is
possible for $\Lambda$ to remain quantum even in contexts where everything
else is deemed classical. 
Finally in Section~\ref{sec:specs} we speculate on cases where 
\textit{ad hoc} Planck parameters scale differently with the length scale
under observation. 

We conclude with a general discussion of the implications of our findings. 

\section{Planck's parameter in standard quantum cosmology}\label{sec:MSS}
Let us consider the minisuperspace (MSS) quantum cosmology (QC) following from the
Einstein-Hilbert action. The reduced action can be written as: 
\begin{equation}\label{Sg}
S=6\kappa V_{c}\int dt\bigg(a^{2}\dot{b}+Na\bigg(b^{2}+k-\frac{\Lambda }{3}%
a^{2}\bigg)\bigg),
\end{equation}
where $\kappa =1/(16\pi G_{N})$ (with $G_N$ Newton's constant), $a$ is the expansion factor, $b$ is the
expansion rate (with $b=\dot{a}/N$ on-shell), $N$ is the lapse function
associated with time coordinate $t$ and $k=0,\pm 1$ is the normalized
spatial curvature~\cite{Vilenkin,MSS}. Crucially, for the purpose of this paper, 
\begin{equation}
V_{c}=\int d^{3}x\sqrt{h},
\end{equation}%
is an integral in comoving spatial variables over the region under study
(assumed to be fixed in comoving variables) and $h$ is the determinant of
the comoving (time-independent) 3-metric. Usually in QC, one sets $k=1$ and
integrates over the whole sphere, so that $V_{c}=2\pi ^{2}$. In some cases,
one considers topologically non-trivial compact spaces with $k=0,-1$. Here,
we propose that the region under study be a generic fixed region in comoving
coordinates, where homogeneity and isotropy can be assumed. Different such
regions may, or not, have different properties, so this partitioning can be
at a scale different from the scale of homogeneity. It can also be a
smoothing scale, if strict homogeneity is not assumed. Whatever the
interpretation, we assume that the different regions do not interact. The
overall wave-function, therefore, is the cross-product of the wave-functions
for different regions, each with an associated $V_{c}$: 
\begin{equation}
\psi =\bigotimes_{i}\psi _{i}(a;k,V_{ci}).
\end{equation}%
Underlying this ansatz is the assumption that some degrees of freedom (such
as the graviton's) are frozen or are ignorable. 
Nevertheless, the integration over the spatial degrees of freedom leaves a trace,
in the form of $V_{ci}$, in the pre-factor of the classical action (\ref{Sg}%
), and in the quantum wave-function for each of these regions. 

In the absence of matter other than $\Lambda $, the pre-factors in (\ref{Sg}%
) are irrelevant for the classical theory. But they are relevant for the
quantum theory, and indeed they contribute to the \textit{effective} $\hbar $%
. They propagate into the Poisson Bracket (PB): 
\begin{equation}\label{PB1}
\{b,a^{2}\}=\frac{1}{6\kappa V_{c}},
\end{equation}%
and upon quantization into the commutator: 
\begin{equation} \label{com1}
\left[ \hat{b},\hat{a^{2}}\right] =\frac{il_{P}^{2}}{3V_{c}}
\end{equation}%
where $l_{P}=\sqrt{8\pi G_{N}\hbar }$ is the reduced Planck length.

Obviously, we could normalize the phase space so that the PB is 1, but if we
insist on working with observables that become our favoured classical
variables \textit{when classical cosmology emerges}, then the function
controlling their commutator is a variable depending on contextual factors,
such as the volume of the region under study, and may vary in time.
For example, we could have worked with a conjugate to $a^2$ given by $\Pi_a^2=6\kappa V_c b$, i.e. a 
version of the Hubble parameter $H$ multiplied by the volume under study. But this  is not 
our ``favoured'' variable: there is no point in 
multiplying what comes out of the Hubble diagram by $V_c$. Within MSS the observable is simply 
$H$ (or $b$), but then the factor of $1/V_c$ on the right hand side (\ref{com1}) is unavoidable.  This factor is not arbitrary. It is determined by the action, and by the choice of relevant observable, which in many cases (including this one) is obvious, 
given the MSS approximation. And yet, this factor has practical consequences, since it affects the quantum fluctuations and uncertainty relations.  It says that these depend on whether we look at the whole Universe, the current Hubble volume, or just the local cluster, as we now explain further.

There is some controversy over the details of the mechanism leading to
classicality (and even the definition of the latter; see, e.g.~\cite{decoh,decoh2,Habib}); but in view of (\ref%
{com1}), whatever brings about the classical limit in quantum cosmology, the
process is controlled by the effective Planck parameter: 
\begin{equation}
\mathfrak{h}=\frac{l_{P}^{2}}{3V_{c}}.  \label{plk}
\end{equation}%
This contains $\hbar $, but also $G_{N}$ and $V_{c}$, so the classical limit
depends on the size of the region under study. Given the presence of $G_N$
(and so $l_P$) 
in the commutator, one might expect that any
region with a physical size larger than the Planck length is classical, but
that is not what the Heisenberg relations tell us.
These can be derived purely kinematically\footnote{%
I.e.: from the commutator alone, without using the Hamiltonian or the
dynamics.}, and result in~\cite{schrod}:
\begin{equation}
\sigma (b)\sigma (a^{2})\geq \frac{\mathfrak{h}}{2},  \label{heis1}
\end{equation}%
from which a dimensionless version can be written: 
\begin{equation}
\frac{\sigma (b)}{\langle b\rangle }\frac{\sigma (a^{2})}{\langle
a^{2}\rangle }\geq \frac{\mathfrak{h}_{1}^{\prime }}{2},  \label{heis1d}
\end{equation}%
with 
\begin{equation}
\mathfrak{h}_{1}^{\prime }=\frac{l_{P}^{2}}{3V_{c}{\langle a^{2}\rangle }{%
\langle b\rangle }}.  \label{plk1}
\end{equation}%
This contextual Planck parameter is dimensionless and depending on whether
it is much larger than 1, or the opposite, the system may be declared
quantum or classical. It depends not only on the comoving volume of the
region under study, but also on time, as the physical size of the region
increases (via the scale factor $\langle a^{2}\rangle $) and the Hubble
scale, implicit in $\langle b\rangle $, decreases. 

For regions and times for which $\sigma (b)\ll \langle b\rangle $ and $%
\sigma (a^{2})\ll {\langle a^{2}\rangle }$, we may approximate 
\begin{equation}
\mathfrak{h}_{1}^{\prime }\approx \frac{l_{P}^{2}}{3{\langle V\rangle }{%
\langle H\rangle }},  \label{plk1approx}
\end{equation}%
where $V$ is the proper volume of the comoving region under study (i.e.:  $%
V=V_{c}a^{3}$) and $H=b/a$ is the Hubble parameter. In this simple model,
the criterion for quantum space-time fluctuations to occur is not that one
looks at length scales smaller than the Planck scale, but smaller than the
scale: 
\begin{equation}
l_{Q}\sim (l_{P}^{2}H^{-1})^{1/3}.  \label{lQ1}
\end{equation}%
This scale increases in time, as $H$ decreases. Nowadays, it is bigger than
the Planck scale, but still microphysical (and hence one may declare the
Friedman metric inapplicable). But taking this calculation as a toy model,
the point remains that the relevant scale is not $10^{19}$~GeV, but around
0.1~GeV. Indeed it is a time-dependent scale, related to the time-dependent
Planck parameter $\mathfrak{h}_{1}^{\prime }$\footnote{After this paper was submitted,
the work of~\cite{Randono} was brought to our attention. The ``mesoscopic'' scale derived
in that reference is the same as ours, for the same reason (which is essentially 
of dimensional analysis). Whilst our result is more general (it does not depend on the 
concrete dynamics) the work of~\cite{Randono}, precisely for being less general,
is more concrete, and even derives wave functions for the Universe incorporating this
scale.}

Note that within the approximation leading to (\ref{plk1approx}) we can also
take the average of the Hamiltonian constraint: 
\begin{equation}
\mathcal{H}=b^{2}+k-\frac{\Lambda }{3}a^{2}=0,  \label{hamconst}
\end{equation}%
associated with some Lagrange multiplier, $N$. Since, in the same
approximation, $\langle b^{2}\rangle \approx \langle b\rangle ^{2}$ and $%
\langle a^{3}\rangle \approx \langle a\rangle ^{3}\approx \langle
a^{2}\rangle ^{3/2}$, this leads to: 
\begin{equation}
\mathfrak{h}_{1}^{\prime }\approx \frac{l_{P}^{2}}{3{\langle V\rangle }\sqrt{%
\frac{\Lambda }{3}-\frac{k}{\langle a\rangle ^{2}}}}.  \label{plk1approx2}
\end{equation}%
Obviously, these assumptions already assume that we are in a regime where $%
\mathfrak{h}_{1}^{\prime }\ll 1$. We could also include a matter term $\rho $
in the Hamiltonian constraint and this would appear in (\ref{plk1approx2}).

\section{Models with a dynamical cosmological constant}\label{sec:GB}

Other models may be considered, for example a version of Einstein-Cartan
theory including a quasi-Euler term, where $\Lambda$ becomes dynamical~\cite%
{alex1,alex2}. The Hubble variable $b$ is then identified with the
parity-even part of the connection, and on-shell $b = \dot a +Ta$,
where $T$ is the parity-even component of the torsion field. A parity-odd
component of the connection and torsion may be present (this is the
so-called Cartan spiral staircase~\cite{spiral}), but we shall set it to
zero here, and confine ourselves to the parity-even branch of the phase
space~\cite{alex1,alex2,MZ}.

Such a theory is interesting for this paper because $\Lambda $, by virtue of
classically being part the phase space, upon quantization becomes an
observable subject to uncertainty relations. Besides the Hamiltonian
constraint, the theory has a constraint forcing the momentum $\Pi $
conjugate to $\Lambda ^{-1}$ to be proportional to the Chern-Simons (CS)
integral. Spelling it out~\cite{MZ}\footnote{%
With some cosmetic modifications with respect to~\cite{MZ}, namely $\Pi
\rightarrow 2\Pi $, and $V\rightarrow 6V$. This prevents the appearance of
effective Planck parameters that differ merely by numerical factors.}, the
action (\ref{Sg}) is extended to: 
\begin{eqnarray}\label{Faction2}
S &=&6\kappa V_{c}\int dt\bigg(a^{2}\dot{b}+\Pi \frac{d\Lambda ^{-1}}{dt} 
\nonumber \\
&&+Na\bigg(b^{2}+k-c^{2}-\frac{\Lambda }{3}a^{2}\bigg)  \nonumber \\
&&+V\bigg(\Pi -(b^{3}+3b(k-c^{2}))\bigg)\bigg),
\end{eqnarray}%
so that the PBs are now (\ref{PB1}) and, 
\begin{equation}\label{PB2}
\{\Lambda ^{-1},\Pi \}=\frac{1}{6\kappa V_{c}},
\end{equation}%
with a new constraint, associated with the Lagrange multiplier $V$, given
by: 
\begin{equation}
\mathcal{V}=\Pi -\tau _{CS}=0  \label{confconst}
\end{equation}%
where : 
\begin{equation}
\tau _{CS}=b^{3}+3bk.
\end{equation}%
It can be shown~\cite{MSS} that this is the imaginary part of the CS
functional in the MSS approximation, stripped of factors arising from the
spatial integration. This quantity has been proposed as a measure of time in
quantum gravity~\cite{chopin}, the so-called Chern-Simons time. 

Since the constraints are first class~\cite{MZ}, the PBs imply the
commutation relation (\ref{com1}) as well as: 
\begin{eqnarray}
\left[\hat \Lambda^{-1},\hat{ \Pi}\right]&=&\frac{ i l_P^2}{3 V_c}=i%
\mathfrak{h }.  \label{com2}
\end{eqnarray}
From these, Heisenberg relations and their effective Planck parameters may
be derived. In addition to (\ref{heis1}), which is still valid, we have: 
\begin{equation}  \label{heis2}
\sigma (\Lambda^{-1}) \sigma (\tau_{CS})\ge \frac{\mathfrak{h }}{2},
\end{equation}
from which a dimensionless version\footnote{We stress that all dimensionless
relations we have written respect the symmetries of the underlying variables. 
For example, if $k=0$, then $a$ can be multiplied by any constant, but this drops out of the dimensionless  
relations. The analogy with $x$ and $p$, where dividing by $\langle x\rangle$ would clash with the translation invariance,
therefore does not apply. None of the variables in our relations has shift invariance.}
can be derived: 
\begin{equation}  \label{heis2d}
\frac{\sigma (\Lambda^{-1})}{\langle \Lambda^{-1} \rangle} \frac{\sigma
(\tau_{CS})}{\langle \tau_{CS}\rangle}\ge \frac{\mathfrak{h}_2^{\prime }}{2}
\end{equation}
with 
\begin{equation}  \label{plk2}
\mathfrak{h}_2 ^{\prime }=\frac{l_P^2}{3V_c {\langle \tau_{CS}\rangle}
\langle \Lambda^{-1}\rangle }.
\end{equation}
This provides us with an example where the same system may have different
effective Planck parameters for different pairs of variables. These may be
wildly different. For example, in the quantum domain (where $\mathfrak{h}%
^{\prime }_i$ are much larger than 1), we could conjure a state with very
different ${\langle a^2 \rangle} {\langle b \rangle}$ and $\langle
\tau_{CS}\rangle\langle \Lambda^{-1} \rangle$, so that $\mathfrak{h}^{\prime
}_1$ and $\mathfrak{h}_2^{\prime }$ differ appreciably. The construction of
coherent states, saturating the bound in such an unusual situation, will
undoubtedly be interesting. One can even envisage hybrid situations, where
the system has gone classical for one pair of variables, while remaining
quantum for another pair. We should stress that most of the conditions for classicality
we wrote down are necessary, but not sufficient conditions.  

In the regime where the fluctuations are small with respect to the averages
(requiring $\mathfrak{h}_{i}^{\prime }\ll 1$), we can take the average of
the Hamiltonian constraint (\ref{hamconst}) and of the new constraint (\ref%
{confconst}), and with simple algebra make further progress. In this
approximation: 
\begin{eqnarray}
\mathfrak{h}_{1}^{\prime } &\approx &\frac{l_{P}^{2}}{3V_{c}a^{3}H} \\
\mathfrak{h}_{2}^{\prime } &\approx &\frac{l_{P}^{2}}{3V_{c}\tau
_{CS}\Lambda ^{-1}},
\end{eqnarray}%
where all the variables on the RHS are averages (corresponding to their
classical values), subject to $H^{2} =\frac{\kappa }{6}\rho +\frac{\Lambda }{3}-\frac{k}{a^{2}} $ and 
$\tau _{CS} =Ha(H^{2}a^{2}+3k)$ 
(this is just a rewrite of Eq.~(\ref{hamconst}) including matter, and of (%
\ref{confconst})). The first equation can be written as: 
\begin{equation}
1=\Omega _{\rho }+\Omega _{k}+\Omega _{\Lambda }
\end{equation}%
with the usual definitions for $\Omega _{i}$, so that: 
\begin{equation}
\frac{\mathfrak{h}_{2}}{\mathfrak{h}_{1}}\approx \frac{3\Omega _{\Lambda }}{%
1-3\Omega _{k}}.
\end{equation}%
Hence, even in the (semi-)classical regime, the two Planck parameters can be
very different: if matter dominates $\Lambda $, for example, or if curvature
is appreciable (such as in the \textquotedblleft crisis\textquotedblright\
scenario of~\cite{silk}, but see~\cite{handley}). It may be possible for $a$ and $b$ to behave
classically for all purposes whereas the fluctuations in $\Lambda$ are still
non-negligible. This could be an interesting interface with the issues
raised by~\cite{unruh}.

\section{More speculative examples}\label{sec:specs}
The examples considered so far were derived from first principles, starting
from a classic dynamics and following standard quantization methods. But we
can be more speculative and entertain more general \textit{ad hoc} Planck
parameters~\cite{QuantConst}. A relevant example for this paper can be found in~\cite{LeeJ}. This differs from the 
cases investigated here so far in two ways. Firstly, an arbitrary
function of $\Lambda $ (not necessarily $\Lambda ^{-1}$, as in Eq.~(\ref%
{com2})) was proposed as the conjugate to the CS time, $\tau _{CS}$%
. Secondly (and more importantly), the Planck parameter postulated in~\cite{LeeJ} does not
depend on $V_{c}$, in contrast with (\ref{plk}). This has
the extreme effect of delocalizing the universe in time 
\textit{on all scales}, when it becomes evident that we live in a sharp
state of the cosmological constant, strongly peaked around zero. Thus, in~%
\cite{LeeJ}, the Planck parameter pertaining to two intensive quantities ($%
\Lambda $ and CS time) is itself an intensive quantity.

We could also consider intermediate situations. 
The reason why a $1/V_{c}$ factor appears in the expressions for $\mathfrak{h%
}^{\prime }$ derived from first principles is that the starting point is an
action which is extensive. Hence the \textit{canonical} conjugate of an
intensive variable in MSS must be extensive, but if we ignore the volume
factor in the resulting observable within MSS (for example caring only about
the Hubble parameter, or the CS time), then the factor of $V_{c}$ appears in
the denominator on the right-hand side of the commutation relations, as we
saw.

This is necessary, if building the quantum theory from the classical one;
but by freeing ourselves from this constraint, we could envisage any other
scaling. If we were to appeal to the holographic principle, for example, the
factor of $V_{c}$ would be the area of the comoving region $A_{c}\sim
l_{c}^{2}$, where $l_{c}$ is the linear dimension of the region under study.
If we were instead to imagine super-extensive scalings, such as the one
proposed in~\cite{BHeleph}, a power of $l_{c}$ higher than 3 would replace $%
V_{c}$: 
\begin{equation}
V_{c}\rightarrow l_{c}^{n}
\end{equation}%
where we can relate $n$ to the exponent $\Delta $ defined in~\cite{BHeleph}
according to $n=3\left( 1+\frac{\Delta }{2}\right) $. If all the time
dependent quantities can be related to the Hubble parameter, we would have a
dimensionless Planck parameter of the form: 
\begin{equation}
\mathfrak{h}^{\prime }\sim \frac{l_{P}^{2}}{l^{n}H^{n-2}}
\end{equation}%
where $l$ is the proper size of the region under examination. Hence the
border between classical and quantum would now be located at scale: 
\begin{equation}
l_{Q}\sim (l_{P}^{2}H^{2-n})^{1/n}.  \label{lQ2}
\end{equation}%
instead of (\ref{lQ1}). The holographic example ($n=2$) leads to the naive
expectation $l_{Q}\sim l_{P}$. The unexpected result that the border between
classical and quantum can be larger than $l_{P}$ results from the
conservative assumption that the conjugate of an intensive quantity must be
extensive. Super-extensive momenta would lead to larger and larger $l_{Q}$.
The situation in~\cite{LeeJ} corresponds to the discontinuous $n\rightarrow \infty $ limit.

As a last speculative example, we may wonder whether our generalized
definition of Planck parameter and of quantum uncertainty may affect
thermodynamical quantities under extreme gravitational situations. For
example, it could be that thermodynamic conjugates, such as the temperature $%
T$ and entropy $S$, are subject to uncertainty relations: 
\begin{equation}
\sigma (S)\sigma (T)\geq \mathfrak{h,}
\end{equation}%
with $\mathfrak{h}$ scaling with the volume and temperature in generic ways.
In quantum gravity motivated situations, we can expect $l_{P}$ to appear in
its expression for $\mathfrak{h}$ but otherwise, it could take any form.

\section{Discussion}

We remark the important anthropic implications of allowing for a 
variable Planck parameter~\cite{Lieb,APbook}. The non-relativistic Schr\"{o}dinger equation for atomic physics possesses a
simple scaling property. By a suitable scaling of variables it can be
written in a form in which no constants of Nature appear~\cite{APbook,HartreeBook}. Considering a situation with 
two different sets of constants (primed and unprimed),
if $E$ was the original energy eigenstate and $E^{\prime }$ is the new
one,  then
\begin{equation}
\frac{\hbar ^{\prime 2}E^{^{\prime }}}{e^{\prime 4}m_{e}^{\prime }\ }=\frac{%
\hbar ^{2}E}{e^{4}m_{e}}.  \label{heqn}
\end{equation}
where $e$ and $m_e$ are the electron's charge and mass. 
Therefore, if an atom exists as a solution of the equation with the unprimed
constants of Nature, then a corresponding atom will exist with constants
given by the primed variables. 

We see that large changes of $\hbar $ into what we think of as the
classical regime still allows atoms to exist: larger $\hbar $ means larger
atoms. All the unusual properties of atomic systems, like water
and DNA, do not depend on these constants if there are simultaneous
variations of the other constants in the scaling. They just depend on the
geometric factors like $2\pi $. 
The uncertainty principle will be made of $\Delta p\sim p\sim m_{e}c$ and $%
\Delta x\sim x$ in the $x$ coordinates, so
\[
\Delta p\Delta x\sim m_{e}cx\gtrsim \hbar 
\]%
whereas in the $^{\prime }$ system we would have $\Delta p^{\prime }\sim
p^{\prime }\sim m_{e}^{\prime }c^{\prime }$ and $\Delta x^{\prime }\sim
x^{\prime }$ so

\[
\Delta p^{\prime }\Delta x^{\prime }\sim m_{e}^{\prime }c^{\prime }x^{\prime
}\gtrsim \hbar ^{\prime }.
\]
For further discussion of the implications the reader is referred to~\cite{APbook}.

To conclude, we have examined situations where a contextual, time and
size-dependent Planck parameter can be defined. At first, we worked in QC
from first principles, deriving such a size and time dependent quantity,
which can also depend on the pair of variables under study. We used this
contextual Planck parameter to define the classical limit of QC, concluding
that, even in the most standard theory, classicality depends on the comoving
region's size, the time and the variables under study. In the standard QC
set up, where one studies a whole closed Universe,
there is a single \textquotedblleft time\textquotedblright\ for the
universe to pop up out of the quantum epoch, possibly after creation \textit{%
ex nihilo}. 
By introducing a space scale into the problem, we have shown that at any
time after the Planck epoch, there are always regions that remain quantum.
These must only be smaller than $l_{P}$ at Planck time: later on, they must
in fact be smaller than $l_{Q}=(l_{P}^{2}H^{-1})^{1/3}$. As $H$ decreases in
time, $l_{Q}$ increases, so that there is always a scale larger than $l_{P}$
where the quantum fluctuations are appreciable.

In theories where the cosmological constant, $\Lambda$, is dynamical it is
possible for the latter to remain quantum even in contexts where everything
else is deemed classical. Could this assist in our understanding of the
cosmological constant problem?

We thank S. Alexander and D. Jennings for discussions related to this paper. JM was
funded by the STFC Consolidated Grant ST/L00044X/1. JDB was funded by STFC

\end{document}